\documentclass{article}
\usepackage{tipa}
\usepackage{spconf,amsmath,graphicx,hyperref}
\usepackage{amssymb}
\usepackage{booktabs}
\usepackage{multirow}

\title{Bridging the Perceptual - Statistical Gap in Dysarthria Assessment: Why Machine Learning Still Falls Short}
\name{Krishna Gurugubelli}
\address{Samsung Research \& Development Institute Bengaluru, India\\krishna.g@samsung.com}

\begin{document}
\ninept
\maketitle
\begin{abstract}
Automated dysarthria detection and severity assessment from speech have attracted significant research attention due to their potential clinical impact. Despite rapid progress in acoustic modeling and deep learning, models still fall short of human expert performance. This manuscript provides a comprehensive analysis of the reasons behind this gap, emphasizing a conceptual divergence we term the ``perceptual-statistical gap''. We detail human expert perceptual processes, survey machine learning representations and methods, review existing literature on feature sets and modeling strategies, and present a theoretical analysis of limits imposed by label noise and inter-rater variability. We further outline practical strategies to narrow the gap, perceptually motivated features, self-supervised pretraining, ASR-informed objectives, multimodal fusion, human-in-the-loop training, and explainability methods. Finally, we propose experimental protocols and evaluation metrics aligned with clinical goals to guide future research toward clinically reliable and interpretable dysarthria assessment tools.
\end{abstract}
\section{Introduction}
Dysarthria comprises a set of motor speech disorders resulting from neurological impairment such as Parkinson's disease, amyotrophic lateral sclerosis (ALS), stroke, or cerebral palsy that affect speech motor control and coordination~\cite{duffy2013motor}. Clinically, dysarthria presents with reduced intelligibility, imprecise articulation, altered prosody, and voice quality changes. Accurate assessment of dysarthria severity and intelligibility is fundamental for diagnosis, monitoring disease progression, and tailoring speech therapy. Currently, speech-language pathologists predominantly use subjective intelligibility tests to evaluate the severity of speech disorders and guide treatment planning~\cite{fox2012intensive,maier2009peaks}. However, subjective assessments are influenced by listener familiarity, contextual cues, and linguistic features, and they can be time-consuming and resource-intensive~\cite{van2009speech,klopfenstein2009interaction}. Objective intelligibility assessment methods, in contrast, are cost-effective, reliable, repeatable, and suitable for remote monitoring. Recent advances suggest that Machine-learning models can predict intelligibility and reveal dysarthria-specific articulatory patterns~\cite{maier2009peaks,constantinescu2010assessing}. 

\begin{table*}[htbp]
\renewcommand{\arraystretch}{2}
\centering
\caption{An overview of dysarthric speech databases. AMSDC: Atlanta motor speech disorders corpus, UA-Speech: Universal access speech, and QoLT: Quality of life technology.}
\begin{tabular}{|l|c|c|c|}
\hline
\multicolumn{1}{|c|}{Database} & Dysarthria & \#Subjects & \multicolumn{1}{c|}{Speech stimuli} \\ \hline
The TORGO database~\cite{rudzicz2012torgo} & \shortstack{Spastic, \& Ataxic}& 8 (5 male and 3 female) & \shortstack{words, sentences, and \\ non-speech sounds}\\ \hline

Nemours database~\cite{menendez1996nemours} & \shortstack{Spastic \& Mixed} & 11 (all subjects are male) & sentences and paragraph \\ \hline

UA-Speech database~\cite{kim2008dysarthric} & Spastic \& Mixed & 16 (12 male and 4 female) & words \\ \hline

New Spanish speech database~\cite{orozco2014new} & Hypokinetic& 50 (25 male and 25 female) & \shortstack{non-speech sounds, \\vowels, words, sentences, \\and paragraph} \\ \hline

\shortstack{Home service dysarthric\\speech database~\cite{nicolao2016framework}}& due to cerebral palsy & 5 (3 male and 2 female) & words \\ \hline

Whitaker database~\cite{deller1993whitaker} & \shortstack{Spastic \& Mixed} & 6 (all subjects are male) & words \\ \hline

AMSDC~\cite{laures2016atlanta} & Spastic, Flaccid, and Mixed & \shortstack{57 (35 male and 22 female)\\ out of 99 subjects have dysarthria}& vowels, words, sentences \\ \hline

\shortstack{QoLT  Korean dysarthric\\speech database~\cite{choi2011design}}& due to cerebral palsy & 100 (65 male and 35 female) & syllables and words \\ \hline

\shortstack{Cantonese dysarthric\\speech corpus~\cite{wong2015development}}& due to cerebral palsy & 11 (6 male and 5 female) & words and short sentences\\ \hline
\end{tabular}
\label{Tab:Databases}
\end{table*}
Automated analysis of pathological speech has a long history. Early works focused on hand-crafted acoustic features with statistical classifiers. As research progressed, researchers incorporated prosodic, spectral, and temporal measures, and later combined these with machine learning models to predict intelligibility and severity. Below we summarize the literature under several themes:
\subsection{Feature families used in dysarthria assessment}
Several studies have investigated feature representations for pathological and dysarthric speech assessment. Kim et al.~\cite{kim2015automatic} analyzed sentence-level variations in prosody, voice quality, and pronunciation, while Rong et al.~\cite{rong2016predicting} and de~la~Torre~\cite{de2002intelligibility} modeled intelligibility as a weighted combination of phonation, articulation, nasality, and prosody features, highlighting the critical role of articulation. Key challenges in pathological speech processing include data sparsity, high-dimensional feature spaces, and reliable feature extraction~\cite{gupta2016pathological}, emphasizing the difficulty of obtaining representations that effectively capture dysarthric speech characteristics. Conventional magnitude spectral features have proven valuable in dysarthria assessment. Formant-based measures, particularly the fundamental frequency (F0) and the second formant, show strong correlation with intelligibility~\cite{wilson2000acoustic,kim2011acoustic}. Auditory-inspired features, derived from models of the middle/external ear and basilar membrane, combined with MFCCs, improve intelligibility assessment~\cite{kadi2016fully}. Excitation source and glottal parameters in time and frequency domains further enhance discrimination between dysarthric and normal speech~\cite{NARENDRA201947,narendradysarthric}. Temporal dynamics, captured through log-energy or modulation spectral representations, also provide important cues for intelligibility~\cite{falk2012characterization}. Perceptual linear prediction (PLP) coefficients and MFCCs have been widely applied for analyzing Parkinsonian dysarthria and assessing severity~\cite{benba2016discriminating, martinez2015intelligibility,kapoor2011parkinson}. Spectral features such as centroid, entropy, flux, asymmetry, slope, kurtosis, and roll-off are effective in characterizing imprecise articulation~\cite{fang2017intelligibility}. Formants and their bandwidths are extensively used to improve dysarthric speech intelligibility~\cite{rudzicz2011acoustic, saranya2012improving, gurugubelli2020duration}, and feature selection studies emphasize that long-term average spectral features~\cite{berisha2013selecting, berisha2014modeling}. Alternative representations, including amplitude and frequency modulation (AM-FM) components, amplitude envelopes, and filterbank-based features, have been explored to capture temporal and spectral patterns characteristic of dysarthria~\cite{ falk2011quantifying, legendre2009discriminating}. These features, particularly long-term temporal envelopes, have shown strong correlation with subjective intelligibility ratings, underscoring their importance for objective dysarthria assessment. Recently, instantaneous spectral features such as, perceptually enhanced single frequency filtering co-efficients (PE-SFCC), and analytic-phase features etc., were explored in dysarthria detection assessment~\cite{chandrashekar2019spectro, gurugubelli2019perceptually, GURUGUBELLI20201}. 
\subsection{Conventional and machine learning models}
In recent years, deep learning methods have demonstrated remarkable potential in dysarthric speech assessment by automatically learning complex patterns from raw speech data~\cite{joshy2022automated, joshy2023dysarthria, joshy2023dysarthria_squeeze}. Convolutional Neural Networks (CNNs) have been extensively applied to process raw waveforms and spectrogram representations, capturing both local and global spectral features that are crucial for evaluating speech intelligibility~\cite{sajiha2024dysarthria, radha2024variable, chandrashekar2019spectro, chandrashekar2020investigation, kodrasi2021temporal}. Meanwhile, Recurrent Neural Networks (RNNs) and their variants, including Long Short-Term Memory (LSTM) networks and Gated Recurrent Units (GRUs), excel at modeling the sequential and temporal dependencies inherent in dysarthric speech, enabling more accurate representation of dynamic speech patterns~\cite{mayle2019diagnosing, bhat2020automatic}. To further enhance performance, hybrid architectures that integrate multiple neural network paradigms such as CNN-LSTM combinations with attention mechanisms have been proposed. These models leverage the complementary strengths of spatial and temporal feature extraction, effectively capturing both fine-grained spectral details and long-range temporal dependencies, resulting in improved intelligibility prediction and robust dysarthria assessment~\cite{korzekwa19_interspeech, millet2019learning, fernandez2020attention, karjigi2023speech}.

Transfer learning has recently emerged as a promising strategy in dysarthric speech assessment, where models pre-trained on large general speech corpora~\cite{javanmardi2024pre, Fritsch2021utterance, bhat2020automatic, rathod23_interspeech, cadet2024study} are fine-tuned on dysarthric datasets~\cite{javanmardi2024exploring} to enhance performance, particularly in scenarios with limited annotated data. This approach leverages knowledge learned from normal speech to improve feature representation and intelligibility prediction in pathological speech. Despite these benefits, transfer learning models can be sensitive to domain shifts between normative and dysarthric speech, often leading to reduced generalization when encountering unseen speakers or speech conditions. Moreover, deep learning models remain highly data-intensive and computationally demanding, which poses practical challenges in resource-constrained clinical environments. To complement end-to-end learning approaches, targeted acoustic and linguistic measures such as Goodness of Pronunciation~\cite{yeo23_interspeech}, vowel space area~\cite{thompson2023vowel}, and phoneme-level articulation metrics~\cite{xue2023assessing} have also been investigated for dysarthric intelligibility assessment. These features provide interpretable insights into speech production deficits and can be integrated with deep learning models to improve both performance and clinical relevance. Recently, text-guided dysarthric speech intelligibility assessment framework that leverages custom keyword spotting\cite{11145948}. Acoustic and linguistic similarities between speech and text representations were explored through cross-attention mechanism in \cite{anuprabha2025multi}. However, these machine learning models are often limited in their ability to efficiently represent long-range dependencies~\cite{shahamiri2023dysarthric}. This limitation is especially critical in dysarthric speech, which is characterized by impaired articulatory control, slowed speech rate, rhythmic disturbances, and intra-speaker variability, all of which make long-term context modeling essential~\cite{falk2012characterization}. 

Despite architectural advances, generalization across datasets and clinical settings remains challenging. Models often capture dataset-specific cues (microphone, speaker identity) rather than pathology, leading to inflated in-dataset performance but poor cross-dataset robustness.

\subsection{Databases for the assessment of dysarthria}
Dysarthric speech databases are important in the automatic detection and assessment of dysarthria. Research areas such as automatic speech recognition, speech synthesis, language identification, speaker recognition, and speaker verification have large resources which allowed to use state-of-art machine learning techniques. On the other hand, machine learning techniques have not been explored much in the dysarthric speech analysis domain due to lack of good resources. The collection of dysarthric speech has been in progress for over two decades. The challenges like pathological speech sub-challenge (Interspeech 2012)~\cite{schuller2012interspeech} and Parkinson's condition sub-challenge (Interspeech 2015)~\cite{schuller2015interspeech} have created the publicly databases which allowed the researchers to address different aspects of pathological speech. The most commonly used databases developed by various research groups for dysarthric speech assessment are listed in Table~\ref{Tab:Databases}.

The datasets includes recordings from a specific microphone used disproportionately for disordered speakers. Models may learn the microphone signature as a proxy for pathology. Moreover, these datasets vary widely in speaker populations, task design (sustained vowels, read text, spontaneous speech), recording conditions, and labeling protocols. Evaluation often reports correlation with expert ratings (Pearson r), mean absolute error (MAE) for severity scores, classification accuracy for binary detection, or ASR WER as an intelligibility proxy. Heterogeneous evaluation practices complicate cross-study comparisons.



\section{Analysis of Human Expert Perception}
Despite decades of research and increasingly advanced deep learning models, dysarthria detection and assessment systems still don’t achieve perfect accuracy. There are several deep and interacting reasons for this, spanning data, human variability, acoustic complexity, and clinical constraints. However, the core reason modern dysarthria assessment systems don’t reach human-level performance is the gap in understanding between human experts and machine learning models. Human expert judgments come from decades of domain knowledge and integrated perception, not raw acoustics alone. This section describes how clinical experts (speech-language pathologists, neurologists) assess dysarthria through multi-level perceptual and contextual reasoning.

\subsection{Cognitive and Perceptual Mechanisms}
Clinicians with expertise in dysarthria assessment rely on sophisticated perceptual and cognitive mechanisms that integrate auditory, linguistic, and motor knowledge. These mechanisms allow them to extract meaningful information from degraded or variable speech signals and to make nuanced judgments about severity and subtype. Key processes include:

Experts can selectively attend to the speech signal amid background noise, reverberation, or competing speakers. They detect salient acoustic cues such as formant transitions, spectral tilt, and temporal envelope modulations, even when the signal is partially degraded\cite{bregman1994auditory}. This selective attention enables clinicians to focus on diagnostically relevant features rather than irrelevant variations in recording conditions or speaker idiosyncrasies. Human listeners leverage their knowledge of language, phonotactics, and lexical probability to predict missing or distorted speech segments~\cite{norris2003perceptual}. This top-down processing allows clinicians to mentally "fill in" unintelligible portions of speech and maintain accurate overall judgments of intelligibility, articulation, and prosody. Such predictive reasoning is crucial when assessing patients with severe dysarthria, where portions of the signal may be ambiguous or absent. Dysarthria manifests across multiple temporal scales, from rapid articulatory gestures to slower prosodic modulations. Experts integrate acoustic information over these varying timescales, enabling the evaluation of micro-articulatory deviations (e.g., subtle consonant distortions) and global speech rhythm or stress patterns~\cite{mesgarani2012selective}. This hierarchical integration supports nuanced assessments that consider both segmental and suprasegmental impairments. Clinicians implicitly reason about the underlying motor mechanisms that produce speech~\cite{locke1980inference}. They infer which articulators tongue, lips, velum, larynx may be impaired and how these deficits manifest acoustically. This articulatory inference allows clinicians to map observed speech deviations onto neuromuscular control issues, bridging perceptual observation and physiological understanding.

\subsection{Clinical Scales and Labeling Practices}
Perceptual ratings by experts are formalized using standardized clinical scales, which serve as the reference or "ground truth" in both clinical practice and research:

Frenchay Dysarthria Assessment (FDA): Evaluates multiple speech subsystems (articulation, resonance, phonation, prosody) and provides both subsystem-specific and global severity ratings. Speech Intelligibility Test (SIT): Focuses on functional speech comprehension and percentage of words correctly understood in controlled tasks. Disease-specific measures: Instruments such as the Unified Parkinson’s Disease Rating Scale (UPDRS) include speech-related items to monitor progression in specific populations~\cite{enderby1980frenchay,movement2003unified}. These scales consolidate perceptual judgments across dimensions and often collapse them into global severity scores. However, inter-rater variability is inherent; even trained clinicians exhibit only moderate agreement for some speech dimensions. This variability imposes a practical ceiling on the accuracy of computational models trained on these labels, highlighting the importance of explicitly modeling label uncertainty in machine learning approaches.

\subsection{Contextual and Compensatory Listening Strategies}
Experienced clinicians use context and adaptive listening strategies to improve the accuracy of their perceptual judgments. Contextual cues such as semantic, syntactic, and pragmatic context helps clinicians disambiguate degraded speech. For example, lexical expectations allow them to infer missing phonemes or syllables. This context-sensitive perception is critical when evaluating highly impaired or irregular speech. Speakers with dysarthria often adopt compensatory articulatory strategies, such as hyperarticulating certain consonants, increasing loudness, or modifying speech rate. Clinicians recognize these adaptations and incorporate them into their severity ratings, differentiating between primary motor deficits and voluntary compensations.

These adaptive strategies are dynamic, influenced by clinical experience, patient history, and the interaction between the speaker and clinician. Such perceptually and cognitively rich evaluations are typically absent in standard machine learning models, which often rely solely on acoustic features without contextual or articulatory inference. This elaboration highlights why human perceptual assessment remains the gold standard in dysarthria evaluation and underscores the perceptual-statistical gap that must be addressed in automated assessment systems.

\section{What Machine Learning Models Learn and Why They Differ}

\subsection{Representations and inductive biases}
Machine learning models encode representations that emerge from a combination of model architecture, training objectives, and the data used for learning. Traditional hand-crafted features, such as Mel-frequency cepstral coefficients (MFCCs) or formants, embed domain knowledge inspired by human auditory processing, providing a structured prior that emphasizes spectral and temporal patterns. In contrast, deep learning models learn hierarchical representations directly from raw input signals, capturing increasingly abstract patterns across layers. Despite these strengths, model inductive biases such as translation invariance in convolutional neural networks (CNNs) or attention patterns in Transformers \textit{do not inherently encode the causal or motoric relationships that underlie speech production}. Consequently, while models can effectively capture statistical regularities in acoustic signals, they may remain insensitive to the articulatory, prosodic, and linguistic cues that clinicians rely on. This mismatch can lead to models predicting surface-level features accurately but may fail in capturing clinically relevant variations tied to neuromotor control.

\subsection{Label-driven learning and the limits of supervision}
Supervised learning frameworks rely on labels typically derived from clinician perceptual ratings of intelligibility or severity as the ground truth. These labels inherently reflect context-dependent judgments, compensatory strategies employed by the speaker, and inter-rater variability. Models trained to minimize numerical differences with such labels therefore learn an “average” mapping, which may not correspond to any single expert’s inference strategy. Furthermore, label noise and limited dataset diversity can exacerbate misalignment. Models may overfit to spurious correlations, such as speaker-specific acoustic idiosyncrasies, microphone characteristics, or environmental artifacts, instead of learning pathology-specific cues. This issue is compounded in small or unbalanced datasets, where statistical regularities unrelated to dysarthria dominate model learning.


\subsection{Theoretical Limits}
To frame a theoretical limits on model performance, consider two sources of irreducible error: (1) noise in expert labels (inter- and intra-rater variability), and (2) the Bayes error given feature representations.

\textbf{Inter-rater agreement and upper bounds on correlation: }If expert labels have limited inter-rater reliability-quantified then any model trained to predict the 'consensus (agreement between labels and features)' cannot surpass this reliability. For regression targets, the maximal achievable Pearson correlation between model output and an individual annotator is bounded by the square root of the annotator's reliability relative to the consensus~\cite{fleiss2013statistical}. This implies a hard ceiling determined by label consistency.   

\textbf{Bayes error and feature insufficiency: } Even with perfect labels, if the features available do not fully separate classes (i.e., distributions overlap), the Bayes error, the minimum achievable classification error given the feature distribution, may be non-zero. In dysarthria assessment, acoustic features can be ambiguous: similar acoustic distortions may arise from different perceptual outcomes, producing irreducible classification or regression error.

These examples demonstrate that conventional supervised models capture statistical regularities rather than the causal and context-aware processes that underlie expert perception. 

\section{Limitations of Current Approaches}
Despite advances in acoustic modeling and machine learning for dysarthria assessment, several critical limitations remain, which motivate the need for more perceptually aligned and clinically robust approaches:
\begin{enumerate}
    \item \textbf{Limited Representation of Human Perception:} Most models rely purely on acoustic features (MFCCs, formants, prosody) or embeddings from deep networks. They do not capture the cognitive and contextual reasoning that human clinicians use, such as semantic predictability, compensatory articulatory strategies, or motor knowledge of speech production. This representational gap leads to systematic misalignment between model predictions and expert judgments.
     
\item \textbf{Inter-rater Variability and Label Noise:} Expert labels, used as ground truth, are inherently variable. Even trained clinicians show only moderate agreement on severity and intelligibility scores. Models trained on these labels inherit the noise, limiting achievable accuracy. This ceiling effect is rarely addressed explicitly in current research, yet it represents a fundamental limit on performance.
\item \textbf{Feature Insufficiency and Overlap:} Acoustic cues alone may be insufficient to fully disambiguate perceptual outcomes. For instance, two speakers can produce acoustically similar speech, yet intelligibility may differ due to prosody or contextual cues. Such feature insufficiency imposes an irreducible error bound on model performance.
\item \textbf{Overfitting to Dataset-Specific Artifacts:} Many models inadvertently rely on spurious correlations, such as microphone type, recording environment, or speaker demographics. This reduces cross-dataset generalization and limits clinical applicability.
\item \textbf{Lack of Multimodal Integration:} Humans assess speech using multiple modalities like auditory, visual (lip/jaw movement), and linguistic context. Most current systems use only audio, missing valuable cues that could improve robustness and alignment with clinical reasoning.
\item \textbf{Limited Explainability and Clinical Interpretability:} Black-box deep learning models often provide severity scores or intelligibility estimates without rationale. Clinicians cannot verify or correct predictions, limiting trust and adoption in real-world settings.
\item \textbf{Evaluation Metrics Not Fully Aligned with Clinical Goals: }Standard metrics (e.g., Pearson correlation, MAE) may not reflect clinically meaningful thresholds, such as the ability to detect functional intelligibility loss or distinguish between mild and severe cases. This misalignment can lead to models that perform well statistically but poorly in practice.
\end{enumerate}


Addressing these limitations is crucial to developing automated dysarthria assessment systems that are not only accurate but clinically meaningful, interpretable, and trustworthy. This article is motivated by the need to move beyond incremental engineering gains (e.g., marginally better classifiers) and instead diagnose the conceptual reasons why models diverge from expert perception. We argue that the root cause is a representational and inferential mismatch: human experts perceive speech as a motor-linguistic communicative act, while machine learning models are optimized to detect statistical regularities in acoustic features. We call this the perceptual-statistical gap. Understanding this gap is crucial for designing algorithms that align with clinical goals.

\section{Bridging the Perceptual-Statistical Gap: Methods and Protocols}
In spite of advances in acoustic modeling and machine learning, current automated dysarthria assessment systems face several limitations that hinder clinical applicability. One key challenge is the representational gap between machine learning models and human perception. Most existing models rely solely on acoustic features, such as MFCCs, formants, prosody, or embeddings from deep networks, and fail to capture the cognitive and contextual reasoning clinicians use when evaluating speech. For example, human experts consider semantic predictability, compensatory articulatory strategies, and motor knowledge of speech production to make judgments about intelligibility and severity. Bridging this perceptual-statistical gap requires integrating perceptually motivated features, human-inspired loss functions, and contextual information that reflect the motor-linguistic nature of speech.

Another critical limitation arises from inter-rater variability and label noise. Even highly trained clinicians demonstrate only moderate agreement when rating severity or intelligibility, and models trained on these labels inherit this inherent variability. This introduces a ceiling effect on achievable accuracy that is rarely addressed in current research. Future work should develop probabilistic or uncertainty-aware models that account for variability in clinician ratings, aggregate multiple annotations to derive consensus labels, and incorporate human-in-the-loop approaches to iteratively refine ground truth data. Feature insufficiency is also a major obstacle. Acoustic cues alone may not fully account for perceptual outcomes, as two speakers may produce acoustically similar speech with differing intelligibility due to prosodic, contextual, or linguistic differences. Addressing this challenge requires multimodal and context-aware modeling that integrates visual articulatory information, linguistic context, and temporal dynamics, better reflecting the cues humans naturally use in speech assessment. Such approaches could improve robustness and reduce the irreducible error inherent in single-modality systems.

Overfitting to dataset-specific artifacts remains a persistent problem. Many models inadvertently learn spurious correlations related to microphone type, recording environment, or speaker demographics, which severely limits cross-dataset generalization and clinical applicability. Research should focus on domain adaptation, data augmentation, and normalization strategies to ensure models generalize across diverse populations and recording conditions, avoiding reliance on irrelevant patterns in the data. Explainability and clinical interpretability are equally important. Black-box deep learning models often provide severity scores or intelligibility estimates without rationale, limiting clinician trust and adoption. Future systems must offer transparent reasoning for predictions, such as feature attributions, visualizations of articulatory deviations, or uncertainty estimates, and integrate clinician feedback to create human-in-the-loop frameworks that are both interpretable and actionable in therapy planning. Finally, evaluation protocols and metrics must be aligned with clinical goals. Standard metrics like mean absolute error or Pearson correlation do not always reflect functionally meaningful differences, such as the ability to detect a clinically significant drop in intelligibility or distinguish between mild and severe dysarthria. Research should develop evaluation frameworks that measure performance against clinically relevant thresholds, assess cross-dataset generalization, and account for longitudinal patient monitoring, ensuring that models are evaluated in terms of practical utility rather than purely statistical performance.

Together, these research directions aim to advance automated dysarthria assessment beyond incremental improvements in classification accuracy toward systems that are robust, perceptually aligned, interpretable, and clinically meaningful. By explicitly addressing the perceptual-statistical gap, integrating multimodal information, accounting for label variability, and aligning evaluation with functional goals, future models can better replicate human judgment and support more effective clinical assessment and management of dysarthria.

\subsection{Proposed Experimental Protocols and Evaluation Metrics}
To rigorously evaluate dysarthria assessment methods intended to bridge the perceptual-statistical gap, we propose standardized experimental protocols:
\begin{enumerate}
    \item Multi-rater annotation: Collect ratings from multiple clinicians for each sample, report inter-rater reliability (ICC) and use consensus or probabilistic labels (e.g., label distributions).
    \item Cross-dataset evaluation: Test models across independent corpora with different recording conditions and languages to assess generalization.
    \item Clinically meaningful metrics: Report correlation with human ratings (Pearson r), but also clinically relevant thresholds (e.g., sensitivity at severity cutoffs), ASR-based intelligibility proxies (WER), and calibration measures.
    \item Data-augmentation: Integration of fairness-aware dysarthric speech augmentation is needed for the optimal performance\cite{m25_interspeech}.
    \item Ablation studies: Evaluate the incremental impact of perceptual features, SSL pretraining, ASR-informed losses, and multimodality.
    \item Explainability evaluation: Assess whether model explanations correspond to clinician judgments using agreement metrics and user studies.
\end{enumerate}

\section{Conclusion}
Humans evaluate speech through meaning, motor control, and context; models assess it through acoustic statistics. The persistent performance gap in dysarthria assessment reflects not just technical limitations but a fundamental divergence in understanding. Until machine learning systems internalize a richer, perception-oriented understanding of speech, their predictions will remain imperfect approximations of human judgment. This manuscript argues that reducing the performance gap between ML models and human experts in dysarthria assessment requires addressing a conceptual mismatch: the perceptual-statistical gap. We reviewed clinical perceptual mechanisms, existing acoustic and ML methods, and theoretical limits due to label noise and feature insufficiency. We proposed concrete strategies-perceptual features, SSL, ASR-informed objectives, multimodal fusion, and human-in-the-loop refinement and experimental protocols to evaluate progress. 

\bibliographystyle{IEEEbib}

\end{document}